\documentclass[conference]{IEEEtran}
\usepackage[utf8]{inputenc}
\usepackage{graphicx}
\usepackage{amsmath}
\usepackage{graphicx}
\usepackage{subcaption}
\usepackage{hyperref}
\usepackage{cite}
\usepackage{url}
\usepackage{booktabs}

\usepackage{etoolbox}
\usepackage{url}
\usepackage{microtype}
\UseRawInputEncoding

\AtBeginEnvironment{thebibliography}{%
  \footnotesize      
  \raggedright       
  \sloppy            
  \setlength{\itemsep}{0.28ex} 
}

\title{LLMs and Agentic AI in Insurance Decision-Making: Opportunities and
Challenges For Africa}

\author{
  \IEEEauthorblockN{%
    Graham Hill\IEEEauthorrefmark{1}, 
    Jing‑Yuan Gong\IEEEauthorrefmark{1}, 
    Thulani Babeli\IEEEauthorrefmark{3}
    Moseli Mots'oehli\IEEEauthorrefmark{2}\IEEEauthorrefmark{1}, 
    James Gachomo Wanjiku\IEEEauthorrefmark{1}, 
  }
  \IEEEauthorblockA{%
   \IEEEauthorrefmark{1}The Shard South Africa; 
    \IEEEauthorrefmark{2}University of Hawai‘i at Manoa; 
    \IEEEauthorrefmark{3}Elenjical Solutions South Africa%
  }
}

\begin{document}

\maketitle

\begin{abstract}
In this work, we highlight the transformative potential of Artificial Intelligence (AI), particularly Large Language Models (LLMs) and agentic AI, in the insurance sector. We consider and emphasize the unique opportunities, challenges, and potential pathways in insurance amid rapid performance improvements, increased open-source access, decreasing deployment costs, and the complexity of LLM or agentic AI frameworks. To bring it closer to home, we identify critical gaps in the African insurance market and highlight key local efforts, players, and partnership opportunities. Finally, we call upon actuaries, insurers, regulators, and tech leaders to a collaborative effort aimed at creating inclusive, sustainable, and equitable AI strategies and solutions: by and for Africans.
\end{abstract}


\section{Introduction}\label{sec:introduction}
Artificial Intelligence (AI), especially Large Language Models (LLMs) and orchestrated agentic AI, is transforming finance and, in particular, how insurers analyze risk, process claims, and make decisions. However, most of the AI action has occurred in the past, outside of African influence, leaving unaddressed gaps and opportunities in African insurance markets. To this end, this paper examines how LLMs and agentic AI can drive value for African insurers by surveying regional research efforts, startups, and established incumbents. We aim to spark a conversation that encourages collaboration among actuaries, insurance professionals, regulators, and tech leaders to build insurance solutions powered by AI, particularly those that are only possible and beneficial for the African market. The remainder of the paper reviews key AI concepts and global use cases, then dives into African initiatives and challenges. Finally, we speak to the rarely discussed aspect of technology adoption: the relationship between new technology and regulation. This is particularly important in highly regulated financial markets like South Africa \cite{mabaso2021twinpeaks}.

\begin{figure}[htbp]
  \centering
    \includegraphics[width=1.01\linewidth]{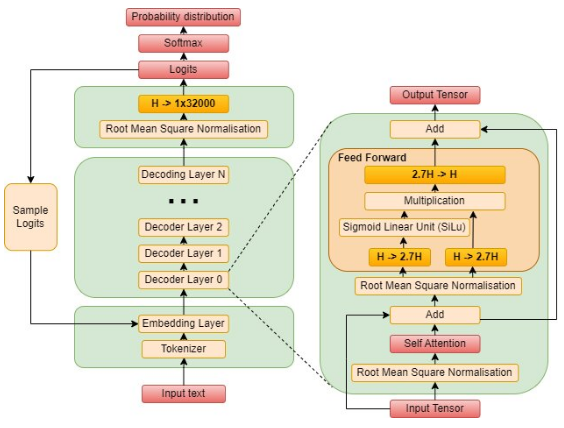}
  \caption{An example transformer LLM (LLaMa2) architecture (Adopted from \cite{Seiranian2024phdthesis}) that processes user prompts end-to-end through stacked self-attention and feed-forward layers.}
  \label{fig:llm}
\end{figure}

\section{LLMs and Agentic AI in Global Insurance}\label{sec:AI_in_insurance}
In this section, we trace the evolution of LLMs from Language Modeling (LM) and Natural Language Processing (NLP), to their role in agentic AI. We state existing global use cases and challenges of LLMs and agentic AI in the sensitive, consequential \cite{luciano2023adversarial}, and highly regulated insurance industry \cite{bis2025regulating}.

\subsection{LLMs and Agentic Frameworks}\label{sec:LLMS_Agents}
Before LLMs, we had smaller LMs \cite{shannon1951prediction}. We made them more complex \cite{Hochreiter1997LSTM,chung2014empirical} and provided them with more data, but they did not perform well and were slow. As computers became faster \cite{owens2008gpu}, we made them bigger \cite{mikolov2010recurrent}, gave them more data \cite{brants2006web} and computational power, and yet hardly moved the needle \cite{jozefowicz2016exploring}. Then, we simplified them \cite{vaswani2017attention} but added even more data \cite{radford2018improving} and computational power \cite{radford2019language}, and finally they worked \cite{brown2020language}. They performed so well at first that we began expecting more from them, but we soon realized they had trouble understanding some difficult concepts \cite{hendryckstest2021} and even some basic ones \cite{bender2021stochastic}. However, with a well-designed prompt, LLMs can provide accurate answers almost instantly, making them ideal for tasks that require quick response \cite{reynolds2021promptprogramming}. 

To date, State of the Art (SOTA) models are multi-modal transformer models \cite{touvron2023llama, openai2024gpt4, anthropic2023claude2, deepseekr1}, trained through schemes like reinforcement learning from human feedback \cite{stiennon2020learning}, mixture of experts\cite{lepikhin2020gshard}, chain of thought fine tuning\cite{wei2022chainofthought}, knowledge distillation \cite{wang2020minilm}, and lately iterative constitutional alignment\cite{chen2024iteralign}. Subsequently, these models are fine‐tuned to specialize in specific knowledge domains: Finance\cite{wu2023bloomberggpt}, insurance\cite{ding2025insqabench}, biology \cite{luo2022biogpt}, law\cite{shu2024lawllm}, and languages: French \cite{martin2020camembert}, Spanish\cite{canete2021spanishbert}, Chinese \cite{devlin2019chinesebert}, and multiple African languages\cite{kossai2022afriberta} that we will examine in Section \ref{sec:African_AI}. Rather than overloading a single LLM, assigning specialized roles to individual models enhances their utility, treating them as agents with distinct tasks and capabilities \cite{yao2023react}, with an orchestrator that chains their outputs \cite{liu2024query} to respond to a user's prompt. Agentic AI frameworks decompose a user's prompt into an explicit planning stage, such that sub-goals are generated and executed by specialized agents, with the response evaluated and refined within the system, all orchestrated by a central controller that iteratively loops through these phases before returning a plausible and possibly correct response to the user. Common frameworks and libraries for agentic AI include LangChain \cite{chase2023langchain}, LlamaIndex \cite{run-llama2023llamaindex}, Microsoft Semantic Kernel \cite{andersen2023semantic}, and Auto-GPT \cite{significant2023autogpt}, all of which provide built-in support for function calls, memory management, and multistep workflow orchestration.

\begin{figure}[htbp]
  \centering
    \includegraphics[width=1.01\linewidth]{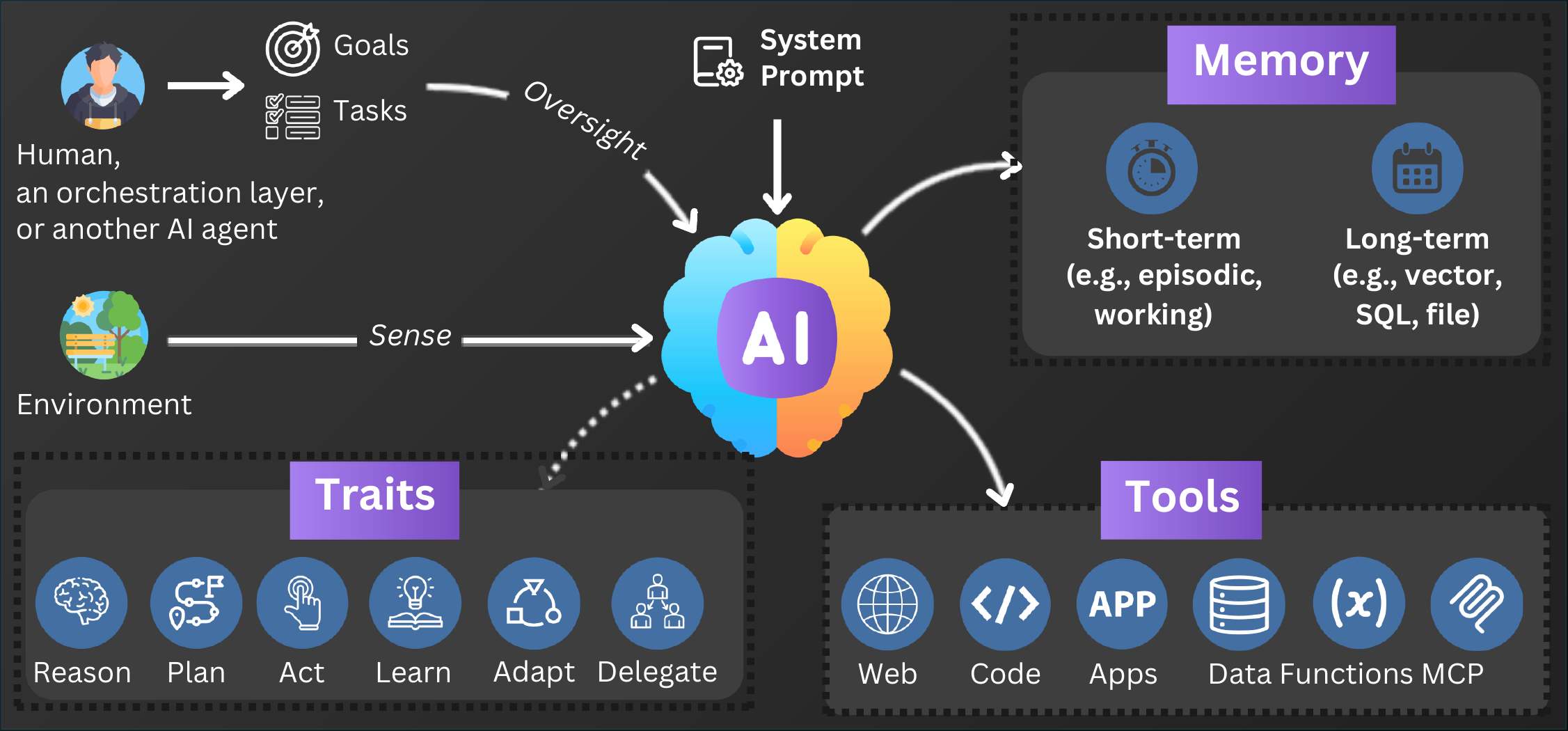}
  \caption{Agentic AI system (Adopted from \cite{huryn2025ultimate}) where a central orchestrator decomposes the request into sub-tasks requiring different tools and memory access, dispatches them to agents with different traits, and recombines their outputs into a coherent, context-aware response.}
  \label{fig:agentic}
\end{figure}

Figures \ref{fig:llm} and \ref{fig:agentic}, respectively, illustrate a transformer block in an LLM mapping prompts to responses, and an agentic system where an orchestrator splits a task into subtasks, dispatches them to specialized agents, and merges their output into context aware responses.

\subsection{Global Use Cases \& Challenges}\label{sec:Global_uses_challenges}
With the increase in open source availability, reasoning capabilities, and utility of LLM, AI agent claims processing solutions have been implemented globally. In Japan, a 100 year old insurance company, \textbf{Fukoku Mutual Life Insurance Company}, replaced as many as 30 claims assessors with the IBM Watson Explorer’s cognitive AI that has natural-language understanding and unstructured data analysis (text, images, audio) capabilities to evaluate medical certificates and calculate claims payouts\cite{mccurry2017japanAI}. Although this deployment preceded the modern LLM era, it delivered a reported 30\% productivity boost and annual savings of ¥140 million while still relying on human sign-off on every claim, highlighting both the promise and limitations of early AI in insurance\cite{mccurry2017japanAI}. \textbf{Tractable.ai} uses AI for vehicle and property damage claims assessments. They primarily rely on Convolutional Neural Networks (CNNs) and not LLMs; their solution is likely upgradable using vision language models to process both damage images and textual descriptions of the car make, model, and other user input information to produce accurate repair estimates. They use Bi-LSTM\cite{Hochreiter1997LSTM} models for Optical Character Recognition (OCR) for automated extraction of critical text from images\cite{orhan2024building}. Tractable.ai partners with GEICO and annually processes over \$2 billion in claims for 35+ insurers, including the largest auto insurers: Tokio Marine and Mitsui Sumitomo (Japan), Covea (France), Aviva (UK), and Admiral Seguros (Spain) \cite{geico_partnership,techcrunch2023tractable}. \textbf{Aidetic}: An Indian AI consulting firm, uses fine-tuned LLMs such as OpenAI's GPT-4 for policy analysis, claims Q\&A, and powering personalized customer support chatbots \cite{aideticChatbots}. As of 2022, Aidetic was reported to be making as much as \$1 million in revenue from their 20+ clients.

\textbf{Shift Technology}, a French AI platform for insurers, combines fine-tuned LLMs and agentic AI to optimize decision-making across underwriting, claims, and fraud detection\cite{shift2014about}. With offices in 10 countries, they serve more than 100 insurers and have analyzed upwards of 5 billion policies, claims, and documents in 2021 alone. Since 2020, the company has integrated LLMs and fine-tuned insurance-specific data to automate document classification, entity extraction, and decision summarization in its claims and fraud workflows. They emphasize model-explainability and improve robustness through human-in-the-loop audit trails to achieve 97\% straight-through claims automation \cite{shift2024claims}. While this is a non-exhaustive coverage, we list and categorize other notable use cases in Table \ref{tab:global_use_cases}

\begin{table*}[htbp]
  \centering
  \begin{tabular}{@{}lllll@{}}
    \toprule
    \textbf{Company} & \textbf{Country} & \textbf{Founded} & \textbf{Task/Category} & \textbf{AI Technology} \\
    \midrule
    \midrule
    \href{https://lelapa.ai}{Lelapa AI}  & South Africa  & 2022  & African language comprehension & Instruction-tuned LLMs  \\
    \href{https://turaco.insure}{Turaco}  & Kenya  & 2018 & Health \& life micro-insurance  & LLM chatbots \& AI validation   \\
    \href{https://naked.insure}{Naked Insurance}  & South Africa  & 2017  & Claims automation   & GPT-3.5 - powered chatbot  \\
    \href{https://pineapple.co.za}{Pineapple}   & South Africa  & 2016  & Claims processing  & CNN vision + LLMs  \\
    \href{https://pulaadvisors.com}{Pula Advisors}  & Kenya  & 2015 & Parametric agri-insurance & Multimodal + LLMs \\
    \midrule
    \href{https://convin.ai}{Convin}  & India & 2020  & Voice-based fraud detection  & Speech-to-text \& LLMs  \\
    \href{https://www.aidetic.in}{Aidetic}   & India  & 2018  & Policy analysis \& chatbots  & Fine‐tuned LLMs  \\
    \href{https://www.claimsforce.com}{Claimsforce} & USA & 2018 & Claims \& workflow automation & AI-assistant \\
    \href{https://www.quantee.com}{Quantee}  & UK & 2017  & Risk \& capital modelling  & Data-driven actuarial models \\
    \href{https://hyperexponential.com}{Hyperexponential} & UK  & 2016  & Pricing \& underwriting optimization & ML-driven pricing \\
    \href{https://www.lemonade.com}{Lemonade}  & USA & 2015  & Claims processing & AI Jim - ML + NLP \\
    \href{https://www.cytora.com}{Cytora} & UK & 2015 & Underwriting \& risk scoring  & LLMs \& ML risk intelligence \\
    \href{https://www.shift-technology.com}{Shift Technology} & France & 2014 & Fraud \& claims automation  & Generative AI \& LLMs \\
    \href{https://tractable.ai}{Tractable} & UK  & 2014 & Damage assessment  & CNN vision + bi-LSTM OCR  \\
    \href{https://www.friss.com}{FRISS (Fraud Risk Intelligence)} & Netherlands & 2012  & Fraud detection  & Generative AI\\
    \href{https://www.insurity.com}{Insurity}  & USA  & 2001  & Policy \& claims management  & Cloud-native AI modules\\
    \href{https://www.scnsoft.com}{ScienceSoft}  & USA  & 1989  & Underwriting \& risk assessment & LLMs \\
    \href{https://www.cccis.com}{CCC Intelligent Solutions} & USA & 1980 & Auto-estimate generation & Computer vision models  \\
    \href{https://www.fukoku-life.co.jp/english/}{Fukoku Mutual Life Insurance }   & Japan & 1923  & Claims assessment & IBM Watson Explorer cognitive AI \\
    \bottomrule
  \end{tabular}
  \caption{Selected insurance-focused AI companies, their origin and founding year, primary task, and the underlying LLM or Agentic AI technology. Lelapa AI notably leads in core multimodal African language tasks in low-resource settings.}
\label{tab:global_use_cases}
\end{table*}

Despite these advances, implementation is not without challenges as insurers need to navigate strict personal data privacy and governance requirements within each jurisdiction. On the technical side, integrating modern AI systems with legacy IT infrastructures\cite{pwc2024ai} remains one of the main bottlenecks, exacerbated by shortages to adequately skilled employees in non-technical companies, and the necessity to manage the substantial costs of custom model development and maintenance\cite{deloitte2025genai}. Despite the potential benefits, gaining stakeholders’ trust and buy-in for these AI systems can be very challenging when you can’t peek under the hood. Without real explainability, transparency, and clear alignment with KPIs, adoption has been seen to be very slow. In the next section, we explore LLM and agentic AI solutions that have been specifically designed or implemented for African languages and cultural contexts.

\section{African AI: Research, Deployments \& Insurance}\label{sec:African_AI}
This section surveys Africa’s AI ecosystem, focusing on LLMs and agentic AI. We review academic and industry research on LLMs and agentic AI across African universities, labs, and startups, then examine their insurance applications

\subsection{African AI Research}\label{sec:AfricanAI}
In South Africa, early deep learning NLP work focused on individual tasks in a single language, for example: \cite{pretorius2020deepmorphSA} for isiZulu morphological analysis and \cite{puoberta2023training} for Setswana masked language modeling, producing embeddings that can then be used to train models for other downstream tasks. In Nigeria, similar masked language models have been built for Yoruba \cite{oseni2020building}, Igbo \cite {eze2022igbobert}, and Hausa \cite{abdullahi2021hausabert}; these models were subsequently fine-tuned for named-entity recognition, sentiment analysis, and question answering tasks \cite{uwa2023nigerianbert}. In Kenya, researchers\cite{ngugi2019bidirectional} have shown the utility of a bidirectional LSTM for the recognition of Swahili named entities. Early works showed we could outperform Google Translate on African languages, but it is initiatives like Masakhane \cite{adelani2021masakhane} through collaborative research that have led to open, curated datasets, benchmarks, and multilingual models of all sizes.

To date, multilingual transformer models such as AfriBERTa \cite{kossai2022afriberta} fine‐tuned on African corpora post competitive results on multiple tasks. In \cite{ojo2023good}, the authors benchmarked GPT\-3.5 and LLaMA variants on eight African languages, showing significant performance degradation in low-resource languages, making the case for custom Afrocentric models. In \cite{prat2024decolonizing}, the authors propose a framework that puts African language and culture at the center when building LLMs. More recent works focus on performance with tiny models with Inkubalm by Lelapa AI, a small language model optimized for a number of low-resource African languages that achieves state-of-the-art results on translation and NER \cite{tonja2024inkubalm}. For a more comprehensive overview of African language modeling efforts, see this survey\cite{alabi2025charting}.

\subsection{Deployments and Gaps}\label{sec:Deployments}
Several insurers across Africa have incorporated AI and LLM technologies into their operations. They include \textbf{Pineapple}, a millennial-friendly insurer \cite{Pineapple2024AIInsurance}, which collaborates with industry leaders like Old Mutual to develop solutions that streamline claims processing. For example, they enable customers to upload images for vehicle claim assessment. In Kenya, \textbf{Pula Advisors} uses AI and remote-sensing data to offer agricultural insurance at scale to over 1.5 million small farmers across 12 countries\cite{techcrunch2021pula}. Their platform uses satellite imagery and predictive weather models to power automated payouts and integrate LLMs into their chat-based customer support tools. Also in Kenya: \textbf{Turaco} offers mobile health and life micro-insurance to more than 3.5 million policyholders\cite{turaco2024website}. They integrate LLMs in both their customer support chatbot and claims validation pipeline, making it possible to achieve a median claims turnaround time as low as 4 hours \cite{turaco2024fastcompany}.

\textbf{Naked Insurance}, based in South Africa, has built a fully digital underwriting and claims engine. They use GPT‑series models to power policyholders facing chatbots, achieving about a 60\% end-to-end claims completion rate. They reportedly settle simple cases in under 15 minutes and have reduced operational costs related to the claims process by nearly 40\% \cite{naked2023claims}. Also in South Africa, \textbf{Lelapa AI}, led by a team of researchers, offers the Vulavula API\cite{tsanni2023lelapa} with five modules, namely: analyze, converse, transcribe, translate, and Speak. These handle audio, text, and conversational tasks in African languages. Their modular approach is well-suited to transform the way different tasks in insurance and finance are done as a whole. The rapid advancement of AI, combined with the growing volume, complexity, and availability of data, has changed the insurance space by enhancing modeling accuracy and efficiency \cite{richman2018ai}, allowing actuaries to build models faster, and allowing actuaries to prioritize tasks that rely on human judgment, which is currently not accurately achievable through AI with no human intervention. These changes are already evident as actuaries adopt LLMs to enrich their workflows, merging structured and unstructured data, automating repetitive processes, and increasing the sophistication of risk models\cite{balona2025operationalizing}. Moreover, other possible uses of LLMs include understanding and providing results for regulatory reports, such as IFRS 17 and other relevant guidelines. By storing the guideline documents in a local search index, an LLM agent can look up the exact sections it needs at query time, then use those passages to craft its answer and cite the specific document or section it referenced, ensuring accuracy and transparency.  Embedding IFRS 17 texts, numerical documents, and PDFs in a vector database enables an LLM agent to automate claims and premium reporting by line of business, answer regulatory queries with precise citations, and generate reports and query‑building code for non‑technical users, offering a practical, automated, and transparent approach to compliance\cite{balona2024actuarygpt}.

\subsection{Opportunities for High-Impact Applications}\label{sec:Opportunities}
Initial LLMs in insurance applied agentic AI to text for bodily harm classification and extent estimation from multilingual accident reports \cite{troxler2024actuarial}, laying the groundwork for actuaries to design personalized products by interpreting complex policy and claim data \cite{balona2024actuarygpt}. Beyond claim processing\cite{dimri2019enhancing,jani2024compliance}, underwriting and pricing\cite{krasheninnikova2019reinforcement,balasubramanian2018insurance}, and mortality modeling \cite{chung2024large}, AI in Africa promises policies based on telematics usage and assistive driving systems that adapt to weather conditions \cite{FAGNANT2015167, MotsOEhli2025Simulating}. These have the potential to remodel personalized risk profiles and can potentially be integrated into agentic AI frameworks. We also envision predictive diagnostics \cite{Jiang2017AIHealthcare}, remote patient monitoring \cite{Topol2019DeepMedicine}, and personalized treatment planning \cite{Reddy2019MachineHealthInsurance} to transform medical insurance through personalized real-time assessments, allowing dynamic premiums and coverage.

\section{Policy, Governance \& Collaborative Pathways}\label{sec:Policy_Governance}
This section discusses the influence of policy, governance, and collaboration on the direction and impact of AI adoption in African insurance.

\subsection{The triple R: Rules, Risks, and Realities}
With the widespread adoption of LLMs across industries \cite{bommasani2021opportunities}, the need for guidelines on their proper integration has gained significant attention from both individuals and governments \cite{gpt4techreport2024}. While regulations governing LLMs exist in some form, their implementation varies: For instance, South Africa has yet to pass a dedicated AI policy (as of this writing). However, existing laws, such as the Protection of Personal Information Act (POPIA) \cite{popia}, impose restrictions on handling personally identifiable information, indirectly affecting LLM deployment \cite{whitecase2024ai}. South Africa and Kenya, two of Africa’s largest economies \cite{statista2024africagdp}, have each proposed AI regulatory frameworks outlining future goals and benefits \cite{kenya2025aistrat,pinsent2024steps}. Understanding these rules is crucial for businesses using LLMs.

This surge in regulatory activity underscores the need for companies, particularly those using or exploring LLMs in their workflows, to carefully consider the implications of handling potentially identifiable information (PII), given its potential impact on job security and the broader economy \cite{jaldi2023artificial}. 
Many companies have adopted LLMs to perform roles in fields such as e-commerce, education, and finance\cite{raza2025industrialllm}, and as such, it is essential to recognize the need to adapt and integrate these systems to benefit business processes across industries. As industries continue to integrate LLMs into critical domains such as finance and insurance, it becomes increasingly clear that regulation alone can not ensure safe, effective, and equitable adoption of AI \cite{morales2023toward}. Addressing challenges such as automated decision-making, data governance, and system interoperability requires a coordinated approach that goes beyond compliance. This underscores the importance of fostering collaborative ecosystems, where regulators, insurers, actuaries, and technology developers work together to co-design inclusive and sustainable AI deployments that align with both ethical and sector-specific realities.

\subsection{Building Inclusive Ecosystems}
The successful integration of LLMs and agentic AI into the financial sector is directly and indirectly tied to the ever-changing policy and regulatory frameworks. POPIA \cite{popia}, for instance, influences how AI is trained and deployed by imposing specific compliance requirements regarding data availability, handling, and storage. As a consequence, regulatory constraints shape not only the technical implementation of LLMs but also establish accountability mechanisms in the event of system failures or misuse. The recent Meta\cite{meta2025lawsuit}, OpenAI, Anthropic\cite{anthropic2025fairuse}, and Google\cite{google2025antitrust} lawsuits demonstrate this and set the future regulatory framework for AI. Integrating advanced AI into rigid, legacy systems: common in highly regulated finance and insurance spaces, requires not only boosting operational efficiency but also ensuring fairness, transparency, and full regulatory compliance \cite{environsciproc2022015066}. This integration demands close collaboration among actuaries, insurers, technology developers, and regulators. Actuaries, as domain experts, need to be trained in the quantification of risks of AI-driven decisions, especially in high‑stakes applications \cite{GALAZ2021101741,gopfert2025risksaidrivenproductdevelopment}. Technology teams also need to build models that are especially interpretable, ethical, and adaptable to evolving regulations. Embedding these goals into the overarching organizational KPIs and building multi-disciplinary teams ensures they are represented and communicated across the enterprise. On the regulatory side, a more proactive approach is required to maintain \cite{fenwick2018business}. By engaging early with industry stakeholders, regulators guarantee practical and effective guidelines that address sector-specific risks without hindering innovation by being risk-averse. It is our strong-held opinion that: Only through the participation of all parties can we deliver AI solutions in the insurance space that are innovative yet locally effective and equitable, an opinion supported in \cite{gracca2021assessment}.

\section{Conclusion \& Call to Action}
In conclusion, this work explores and brings to light existing efforts to integrate LLMs and agentic AI in the insurance industry, particularly in Africa. We highlight how these technologies are reshaping workflows, regulatory thinking, and customer experiences across global markets. We showcase African contributions that are often overlooked in a fast-paced field, and make the case that the “home game” is how we win: African researchers, insurers, developers, regulators and most importantly, the insured, have the opportunity not only to adopt global solutions but to build uniquely local ones. We come up with a simple and clear call to action: We need Africans to think about, bring to existence, and maintain AI-based insurance solutions in the African market. To state the obvious, why this should be the case: your model is generally as good as your data, and we generate the data, live the realities behind it, own and understand it better than anyone.

\bibliographystyle{IEEEtran}
\bibliography{main}
\end{document}